\documentclass[aps,prl,twocolumn,superscriptaddress]{revtex4-1}
\usepackage{graphicx}
\usepackage{epsfig}
\usepackage{amsmath,amsfonts,amsthm}
\usepackage{amssymb}
\usepackage{braket}
\usepackage{units}
\usepackage{multirow}
\usepackage{mathrsfs}
\usepackage{theorem}
\usepackage{bm}
\usepackage{bbold}
\usepackage{threeparttable}
\usepackage{url}
\usepackage[T1]{fontenc}
\usepackage{csquotes}
\usepackage{natbib}
\usepackage{float}
\usepackage{appendix}
\usepackage[colorlinks=true,linkcolor=blue,citecolor=blue,urlcolor=blue]{hyperref}
\usepackage{bookmark}
\usepackage{soul}
\MakeOuterQuote{"}
\usepackage{dcolumn}

\newcommand{\beq}{\begin{equation}}
\newcommand{\eeq}{\end{equation}}
\newcommand{\bqa}{\begin{eqnarray}}
\newcommand{\eqa}{\end{eqnarray}}
\newcommand{\nn}{\nonumber}

\newcommand{\rt}[1]{\sqrt{#1}\,}

\newcommand{\sq}[1]{\left[ {#1} \right]}

\newcommand{\an}[1]{\left\langle{#1}\right\rangle}
\newcommand{\tr}[1]{{\rm Tr}\sq{ {#1} }}
\begin{document}

\title{Experimental Measurement-Device-Independent Quantum Steering and Randomness Generation Beyond Qubits}

\author{Yu Guo}
\affiliation{CAS Key Laboratory of Quantum Information, University of Science and Technology of China, Hefei, 230026, People's Republic of China.}
\affiliation{CAS Center For Excellence in Quantum Information and Quantum Physics, University of Science and Technology of China, Hefei 230026, P.R. China.}

\author{Shuming Cheng}
\email{drshuming.cheng@gmail.com}
\affiliation{Centre for Quantum Dynamics, Griffith University, Brisbane, QLD 4111, Australia.}
\affiliation{CAS Key Laboratory of Quantum Information, University of Science and Technology of China, Hefei, 230026, People's Republic of China.}
\affiliation{CAS Center For Excellence in Quantum Information and Quantum Physics, University of Science and Technology of China, Hefei 230026, P.R. China.}

\author{Xiaomin Hu}
\affiliation{CAS Key Laboratory of Quantum Information, University of Science and Technology of China, Hefei, 230026, People's Republic of China.}
\affiliation{CAS Center For Excellence in Quantum Information and Quantum Physics, University of Science and Technology of China, Hefei 230026, P.R. China.}

\author{Bi-Heng Liu}
\email{bhliu@ustc.edu.cn}
\affiliation{CAS Key Laboratory of Quantum Information, University of Science and Technology of China, Hefei, 230026, People's Republic of China.}
\affiliation{CAS Center For Excellence in Quantum Information and Quantum Physics, University of Science and Technology of China, Hefei 230026, People's Republic of China.}

\author{En-Ming Huang}
\affiliation{CAS Key Laboratory of Quantum Information, University of Science and Technology of China, Hefei, 230026, People's Republic of China.}
\affiliation{CAS Center For Excellence in Quantum Information and Quantum Physics, University of Science and Technology of China, Hefei 230026, People's Republic of China.}

\author{Yun-Feng Huang}
\affiliation{CAS Key Laboratory of Quantum Information, University of Science and Technology of China, Hefei, 230026, People's Republic of China.}
\affiliation{CAS Center For Excellence in Quantum Information and Quantum Physics, University of Science and Technology of China, Hefei 230026, People's Republic of China.}

\author{Chuan-Feng Li}
\email{cfli@ustc.edu.cn}
\affiliation{CAS Key Laboratory of Quantum Information, University of Science and Technology of China, Hefei, 230026, People's Republic of China.}
\affiliation{CAS Center For Excellence in Quantum Information and Quantum Physics, University of Science and Technology of China, Hefei 230026, People's Republic of China.}

\author{Guang-Can Guo}
\affiliation{CAS Key Laboratory of Quantum Information, University of Science and Technology of China, Hefei, 230026, People's Republic of China.}
\affiliation{CAS Center For Excellence in Quantum Information and Quantum Physics, University of Science and Technology of China, Hefei 230026, People's Republic of China.}

\author{Eric G. Cavalcanti}
\email{e.cavalcanti@griffith.edu.au}
\affiliation{Centre for Quantum Computation and Communication Technology (Australian Research Council),
Centre for Quantum Dynamics, Griffith University, Gold Coast, QLD 4222, Australia.}

\begin{abstract}
In a measurement-device-independent or quantum-refereed protocol, a referee can verify whether two parties share entanglement or Einstein-Podolsky-Rosen (EPR) steering without the need to trust either of the parties or their devices. The need for trusting a party is substituted by a quantum channel between the referee and that party, through which the referee encodes the measurements to be performed on that party's subsystem in a set of nonorthogonal quantum states. In this Letter, an EPR-steering inequality is adapted as a quantum-refereed EPR-steering witness, and the trust-free experimental verification of higher dimensional quantum steering is reported via preparing a class of entangled photonic qutrits. Further, with two measurement settings, we extract $1.106\pm0.023$ bits of private randomness per every photon pair from our observed data, which surpasses the one-bit limit for projective measurements performed on qubit systems. Our results advance research on quantum information processing tasks beyond qubits.
\end{abstract}

\date{\today}% It is always \today, today,

\maketitle

{\it Introduction.---}Einstein-Podolsky-Rosen (EPR) steering~\cite{Wiseman2007, Jones2007, Cavalcanti2009} is a class of nonlocal quantum correlations strictly intermediate between entanglement and Bell nonlocality~\cite{Wiseman2007,Quintino2015}: some entangled states are not steerable, and some steerable states are Bell local. It has found applications in information-processing tasks such as one-sided device-independent QKD~\cite{Branciard2012}, subchannel discrimination~\cite{Piani2015,Sun2018}, and randomness generation~\cite{Law2014,Passaro2015,Skrzypczyk2018,Wang2018}.

Entanglement, EPR steering, and Bell nonlocality can be interpreted as the task of entanglement verification with varying levels of trust~\cite{Wiseman2007,Jones2007}, where a referee, Charlie, wants to certify that two parties, Alice and Bob, share entanglement. If Charlie trusts both Alice and Bob (and their devices), it is sufficient for them to violate an entanglement witness. If Charlie trusts neither of them, entanglement can be verified only by violating a Bell inequality. If Charlie trusts one of them but not the other, they need to violate an EPR-steering inequality~\cite{Cavalcanti2009}. Several experiments have been reported to witness EPR steering for qubits~\cite{Saunders2010,Bennet2012,Smith2012}, high-dimensional systems beyond qubits~\cite{Zeng2018}, and continuous variables~\cite{Handchen2012}.

In a seminal work~\cite{Buscemi2012}, Buscemi showed that by equipping Charlie with quantum channels to Alice and Bob, entanglement can be certified for all entangled states in a measurement-device-independent (MDI) way---i.e.~even when Charlie does not trust Alice and Bob. This was further explored in~\cite{Cavalcanti2013,Branciard2013,Xu2014,Verbanis2016,Shahandeh2017} and extended by Cavalcanti {\it et al.}~\cite{Cavalcanti2013} to the case of EPR steering: With access to a quantum channel to Bob and a classical channel to Alice, Charlie can certify entanglement for all EPR-steerable (from Alice to Bob) states.  An experimental MDI verification of steering for qubits was reported in~\cite{Kocsis2015}, together with a method to construct quantum-refereed steering (QRS) witnesses from a given steering inequality. Further discussion of this case was also given in Ref.~\cite{Hall2016} and its quantification in REf.~\cite{HuanYu2018}. In parallel with these developments, growing interest has been devoted to high-dimensional (HD) entanglement because of its potential to provide higher channel capacity~\cite{Liu2002,Grudka2002,Hu2018}, noise robustness~\cite{Vertesi2010,Zeng2018}, and advantages in QKD~\cite{Cerf2002,Groblacher2006,Sheridan2010}. 

Here, we study the trust-free verification of EPR steering beyond qubits. First, an experimentally friendly QRS witness is constructed from a steering inequality, and a specific steering inequality with two measurement settings is programmed into the MDI scenario for qudits. Then, we report the first MDI verification of quantum steering of qudits by generating a class of photonic qutrit pairs.  We then apply our observed data to extract as much as $1.106\pm 0.023$ bits of private randomness per entangled pair, observing the first MDI random number generator that beats the bound of 1 bit for qubit systems with projective measurements~\cite{antonio2016,gomez2018}.

{\it Preliminaries.---}In an entanglement verification protocol, Charlie communicates the measurements to be performed by Alice and Bob on their respective subsystems via labels $x$ and $y$. Alice and Bob respond with their respective measurement outcomes $a$ and $b$. Assuming that the experimental runs are interchangeable, the data collected by Charlie in this experiment are completely encoded by the probability distribution $p(a,b|x,y)$. In the EPR-steering scenario, Charlie will be convinced that they share entanglement, or equivalently, the data demonstrate EPR steering, if and only if the measurement statistics $p(a,b|x,y)$ cannot be described by a Local Hidden State model (LHS)~\cite{Wiseman2007}, i.e.~a model of the form:
\begin{eqnarray}
p(a,b|x,y)  =  \sum_\lambda p(\lambda)\, p(a|x,\lambda)\, \mathrm{Tr}[E^B_{b|y} \rho_{B}^\lambda],  \label{eq:LHSmodel}
\end{eqnarray}
where $\rho_B^\lambda$ is a local quantum state for Bob's system, classically correlated with Alice's system via a random variable $\lambda$ that specifies some arbitrary probability distribution $p(a|x,\lambda)$ for Alice. As Bob and his device are trusted, the probability $p(b| y, \lambda)$ of his measurement outcome can be calculated by the quantum probability rule via an element $E^B_{b|y}$ from some positive-operator-valued measure (POVM) $\{E^B_{b|y}\}_b$  (satisfying $E^B_{b|j} \geq 0$ and $\sum_bE^B_{b|j}=\mathbb{I}^B$) acting on $\rho_B^\lambda $. EPR steering can be detected via the violation of a linear EPR steering inequality~\cite{Cavalcanti2009} of the form
\begin{eqnarray}
W_{\rm S} = \sum^k_{j=1} \an {a_j \hat{B}_j} \equiv \sum_{b,j} g_{b,j} \an{a_j E^B_{b|j}}\leq 0 \;, \label{steerineq}
\end{eqnarray}
where each term for $j\in\{1,\dots, k\}$ is a correlation between the outcomes $a_j$ of Alice's measurement $x=j$ and a quantum observable $\hat{B}_j$ (corresponding to $y=j$) for Bob. The latter can be decomposed in terms of a POVM (or orthogonal projectors as a special case) as $\hat{B}_j \equiv \sum_b g_{b,j} E^B_{b|j}$, with real coefficients $g_{b,j}$. Although other forms of steering inequalities have also been proposed~\cite{Cavalcanti2009,Schneeloch2013,Cavalcanti2015,Zhu2016}, given a quantum state, an optimal linear $W_S$ can be found via a semidefinite program~\cite{Cavalcanti2017}.

For example, consider a scenario with two measurements per party, and define a steering parameter
\begin{equation}\label{eq:SI}
S=\sum_{a=b} p(a,b|j=1) + \sum_{a+b=0} p(a,b|j=2)  \, ,
\end{equation}
where $b$ denotes the outcomes of two mutually unbiased measurements $\hat{B}_j$, $j\in\{1,2\}$, on the $d$-dimensional system $B$, and $a+b$ denotes sum modulo $d$. It is easy to check that $S$ is upper bounded by $2$ and saturates this bound with appropriate measurements acting on the maximally entangled state $|\Phi_d\rangle= \frac{1}{\sqrt{d}}\sum_{i=0}^{d-1}  | i i\rangle$~\cite{SM}. Further, we can show that if there is a LHS model for all $p(a, b|j)$ as in Eq.~(\ref{eq:LHSmodel}), then $S$ is upper bounded by~\cite{SM}
\beq
S\leq S_{\rm LHS}\equiv   1+\frac{1}{\rt{d}}. \label{steerqudit}
\eeq
Hence, a steering inequality of form~(\ref{steerineq}) for qudits can be constructed as $W_{\rm S}=S-S_{\rm LHS}\leq 0$. Note that it is similar to the temporal steering inequality derived in Refs.~\cite{Li2015,Zeng2018}.

{\it Quantum-refereed steering witnesses.---}In the framework of MDI verification of quantum steering, Charlie has the ability to encode Bob's questions in a set of nonorthogonal quantum states rather than classical questions. This gives Charlie the ability to verify entanglement for all EPR-steerable states even without trusting Bob to perform the POVMs $\{E^B_{b|y}\}_{b,y}$ as per Eqs.~(\ref{eq:LHSmodel}) and~(\ref{steerineq}). Instead, these measurements can be programmed into the ``question-states,'' chosen so as to model Charlie asking Bob the following question: ``If you were to perform measurement $y=j$, would you get outcome $b$?''

Specifically, these states are chosen as density matrices $\{\tau^T_{b,j}\}_{b,j}$ on a Hilbert space $\mathcal{H}_C$ of dimension $d$ equal to that of Bob's local state space $\mathcal{H}_B$~\cite{Branciard2013,Hall2016} ($T$ describes the transpose operation in basis $\{|i\rangle\}$), such that $E^B_{b|j}=\tau_{b,j}$. Bob's answer can always be modeled via a POVM $\{\mathcal{B}_1, \mathbb{I}-\mathcal{B}_1\}$ acting on $\mathcal{H}_B \otimes \mathcal{H}_C$, where $\mathcal{B}_1 (= | \Phi_d\rangle\langle \Phi_d|$ in Bob's optimal strategy) models the answer ``Yes''. Denoting by $P(a,{\rm Yes}\,|b,j)$ the probability that Alice answers $a$ and Bob answers ``Yes'' when Charlie asks questions $x=j$ and $\tau^T_{b,j}$, this yields a quantum-refereed steering witness~\cite{SM}
\begin{align}
W_{\rm QRS}=\sum_{a,b,j} g_{b,j} a_j P(a, {\rm Yes}\,|j,\tau^T_{b,j})=\frac{1}{d} W_{\rm S} \leq 0 \; . \label{QRS}
\end{align}
If a LHS model can reproduce the data $P(a, {\rm Yes}\,|b, j)=p(a, b|j)/d$, then $W_{\rm QRS}$ is never positive; the violation of the above inequality, equivalent to violating a standard steering inequality~(\ref{steerineq}), witnesses EPR steering in a MDI way~\cite{SM}. Combining the steering inequality~(\ref{steerqudit}) with the method above produces a QRS witness, which will be experimentally tested with a pair of entangled photonic qutrits. 

%%%%%%%%%%%%%%%%%%%%%%%%%%%%%%%%%%%%%%%%%%%%%%%%%%%%%%%%%%%%%%%%%%%%%%%%%%%%%%%%%%
\begin{figure*}[htbp]
	%[tbph]
\begin{center}
\includegraphics [width= 1.8\columnwidth]{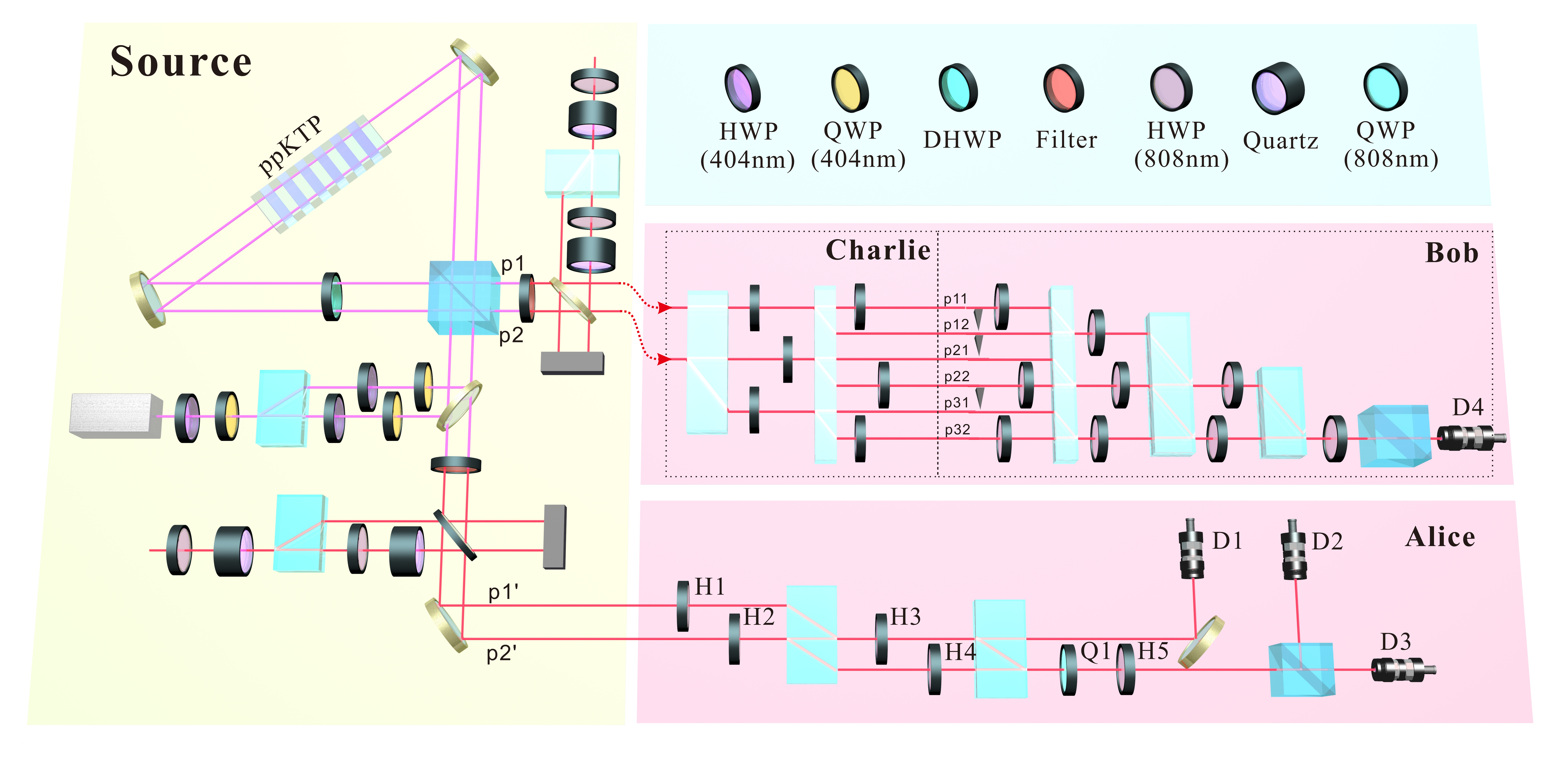}
\end{center}
\caption{Experimental setup for the MDI verification of quantum steering and randomness generation. It consists of three parts: The source (yellow region) illustrates the preparation of the class of states as per Eq.~(\ref{mixedstate}); the pink region named Alice realizes two measurements given in Eqs.~(\ref{b1}) and (\ref{b2}) on Alice's subsystem; the pink region named Bob and Charlie generates the question states $\{\tau^T_{b,j}\}$ sent from Charlie and the partial Bell state measurement. In particular, a class of entangled two-qutrit states is encoded in the hybrid of the path and polarization degrees of freedom of photons, and noise is added with a pair of coherence-destroyed photons. Alice's measurements are implemented via the configuration composed of a series of HWPs, BDs, QWP, and PBS. The question states are encoded in the extra path degree of freedom of the photon, and the measurement projector onto $|\Phi_3\rangle$ is realized in the same way as the state preparation process. BD: beam displacer; PBS: polarizing beam splitter; HWP: half-wave plate; DHWP: dichroic half-wave plate; QWP: quart-wave plate; \emph{D}: single photon detector.}\label{fig:1}
\end{figure*}
%%%%%%%%%%%%%%%%%%%%%%%%%%%%%%%%%%%%%%%%%%%%%%%%%%%%%%%%%%%%%%%%%%%%%%%%%%%%%%%%%

{\it The steering inequality and maximal randomness generation.---}Next, we study the amount of randomness that can be certified from our results, in terms of the optimal probability $P_{\mathrm{guess}}(x^*)$ for an eavesdropper, Eve, to guess the outcome of Alice's result for a given measurement setting $x=x^*$. It was shown in~\cite{Skrzypczyk2018} that the maximal amount of randomness that can be certified via quantum steering, as quantified by the min-entropy
\begin{equation}\label{eq:RG}
H_{min}(x^*)=-\mathrm{log}\, P_{\mathrm{guess}}(x^*) \, , 
\end{equation}
is $\log d$ for qudits with two measurement settings. It is found that given the steering inequality~(\ref{steerqudit}), this maximum can, in principle, be obtained for both of Alice's measurement settings on the entangled state $|\Phi_d\rangle$. We will verify the randomness generation from the observed data in the experiment for $d=3$.

{\it Experimental setup.---}The experimental setup to implement the trust-free verification of high-dimensional quantum steering and randomness generation is given in Fig.~\ref{fig:1}. It is decomposed into three parts: the state preparation, simulation of Alice's measurements, and realization of the input states sent from Charlie and Bob's generalized partial Bell state measurement (BSM).

In the state preparation process, a continuous-wave violet laser at 404 nm is used to generate a pair of entangled photons via a type-II phase-matched spontaneous parametric down-conversion (SPDC) process in a Sagnac structure. The path and polarization degrees of freedom of photons are encoded as the desired states beyond the qubit state space. In particular, as shown in the yellow region of Fig.~\ref{fig:1}, the vertical-polarization photon passing path $ p_1$ or $p_1'$ encodes state $|0\rangle$, and the horizontal-polarization photon in the path $p_2$ or $p_2'$ encodes state $|1\rangle$, while state $|2\rangle$ is for the vertically polarized photon going through $p_2$ or $p_2'$. Hence, the SPDC process yields
$\alpha_0|0\rangle+\alpha_1\mathrm{e}^{i\varphi_1}|1\rangle+\alpha_2\mathrm{e}^{i\varphi_2}|2\rangle \xrightarrow{\rm SPDC}
\alpha_0|00\rangle+\alpha_1\mathrm{e}^{i\varphi_1}|11\rangle+\alpha_2\mathrm{e}^{i\varphi_2}|22\rangle,$ where real coefficients $\alpha_s$ and $\varphi_s$ with $s=0, 1, 2$ depend on the varying angles of the half- and quarter-wave plates (HWPs and QWPs) at 404 nm. In this experiment, we prepare a 3-dimensional isotropic state
\beq
\rho=p|\Phi_3\rangle\langle\Phi_3|+\frac{1-p}{9}\mathbb{I},~~~p\in [0.6, 1], \label{mixedstate}
\eeq
where $|\Phi_3\rangle$ is the maximally entangled state $\frac{1}{\sqrt{3}}(|00\rangle+|11\rangle+|22\rangle)$, and the white noise is added by inserting quartz crystals to completely destroy the coherence of a photon-pair state $\frac{1}{3}(|0\rangle+|1\rangle+|2\rangle)\otimes(|0\rangle+|1\rangle+|2\rangle)$~\cite{Verbanis2016}.

%%%%%%%%%%%%%%%%%%%%%%%%%%%%%%%%%%%%%%%%%%%%%%%%%%%%%%%%%%%%%%%%%%%%%%%%%%%%%%%%%%
\begin{figure}[tbph]
	\includegraphics [width=1.05\columnwidth]{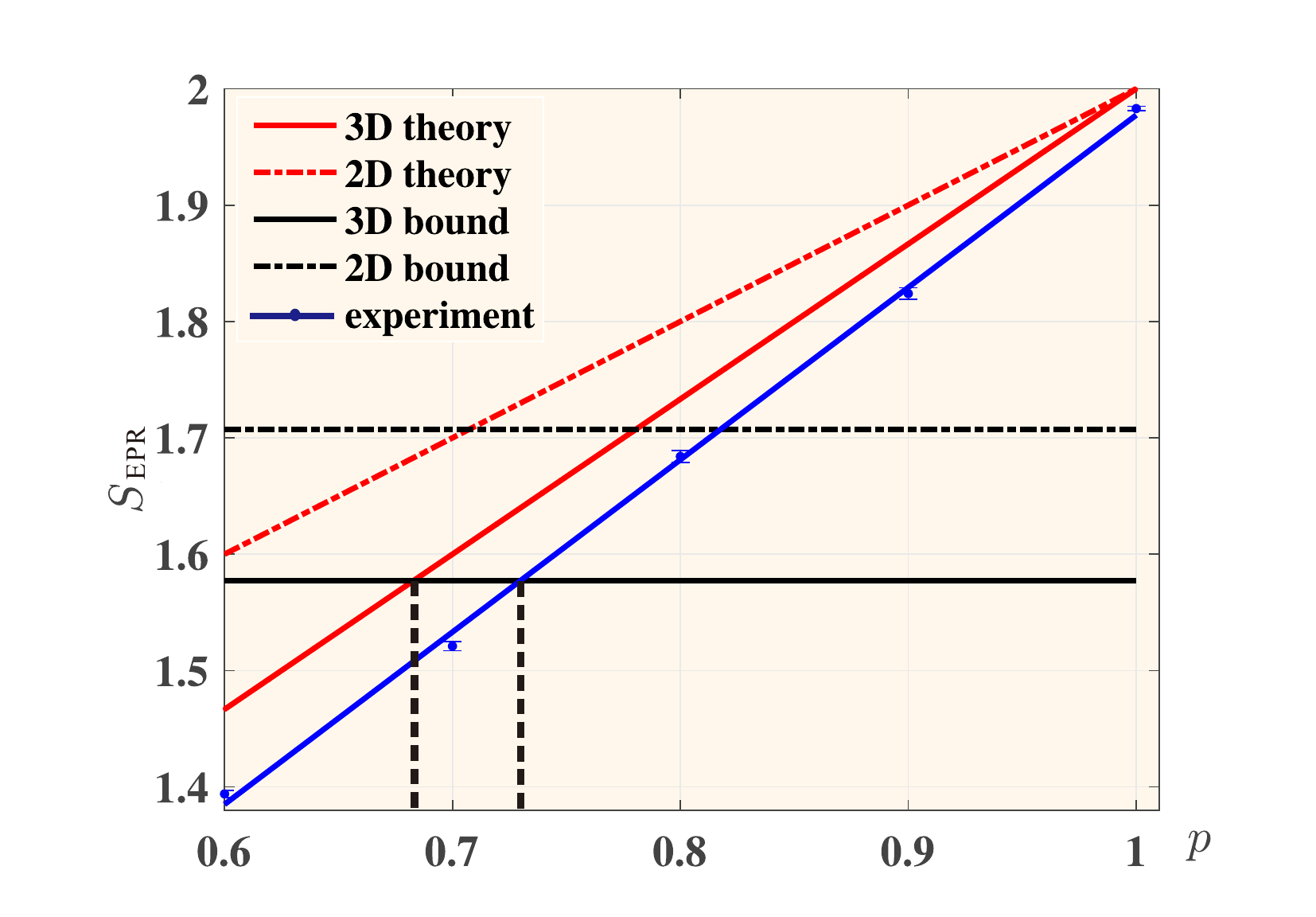}
	\caption{Steering parameter $S$ with $p \in [0.6, 1]$. Our experimental data for the class of states~(\ref{mixedstate}) are marked as blue points, and then fitted in the solid blue line. The theoretical prediction of $S_{\rm EPR}$ for $d=3$ is plotted in the solid red line, while the corresponding bound $S_{\rm LHS}$ for LHS models is shown in the horizontal black line. For contrast, we also plot the theoretical predictions of $S_{\rm EPR}$ (dotted red line) and $S_{\rm LHS}$ (dotted black line) for $d=2$. Note that the minimum value of $p$ for Alice to demonstrate steerability is $0.683$ for the theoretical prediction and $0.730$ for our fitted data, which are highlighted by two dotted black vertical lines. The error bars are of the order of $10^{-3}$, much smaller than the marker size.  }
	\label{fig:2}
\end{figure}

%%%%%%%%%%%%%%%%%%%%%%%%%%%%%%%%%%%%%%%%%%%%%%%%%%%%%%%%%%%%%%%%%%%%%%%%%%%%%%%%%

With respect to the MDI verification of steering as per Eq.~(\ref{steerqudit}), Alice can randomly perform two measurements on her qudit. For qutrits, three measurement outcomes of the setting $j=1$ admit a quantum-mechanical description with the state vectors
\begin{eqnarray}
|b_1=0\rangle = |0\rangle,\, |b_1=1\rangle = |1\rangle, \, |b_1=2\rangle = |2\rangle, \label{b1}
\end{eqnarray}
while the outcomes of the setting $j=2$ which is chosen as a mutually unbiased basis of $j=1$ have a quantum realization as
\begin{eqnarray}
|b_2=0\rangle & = & \frac{1}{\sqrt{3}} (|0\rangle + |1\rangle + |2\rangle), \nn \\
|b_2=1\rangle & = & \frac{1}{\sqrt{3}} (|0\rangle + \mathrm{e}^{i 2\pi/3}|1\rangle + \mathrm{e}^{i 4\pi/3} |2\rangle), \nn \\
|b_2=2\rangle & = & \frac{1}{\sqrt{3}} (|0\rangle + \mathrm{e}^{i 4\pi/3}|1\rangle + \mathrm{e}^{i 8\pi/3} |2\rangle). \label{b2}
\end{eqnarray}
As depicted in the pink region named Alice in Fig.~\ref{fig:1}, these two settings on Alice's qutrit are realized via placing five HWPs, a QWP, two beam displacers (BDs), a polarization beam splitter (PBS), and the single photon detectors sequentially. Specifically, tuning the QWP at $0^\circ$, we rotate the HWP1-5 at $45^\circ$, $0^\circ$, $45^\circ$, $45^\circ$, $0^\circ$ for $j=1$, and set HWP1-5 at $45^\circ$, $67.5^\circ$, $72.37^\circ$, $45^\circ$, $22.5^\circ$ for $j=2$. The detectors \emph{D}1-\emph{D}3 are used to record three outcomes $0-2$, respectively.

The third part in the pink region named Bob and Charlie of Fig.~\ref{fig:1} shows the realization of the question states $\{\tau^T_{b,j} \}$ sent from Charlie and the implementation of the partial three-dimensional BSM $\{\mathcal{B}_1, \mathbb{I}-\mathcal{B}_1 \}$ on Bob's distributed qutrit and the received states. First, these question states are encoded on the extra dimensions of the path degree of freedom of photons, instead of an auxiliary particle~\cite{Xu2014,Kocsis2015}, and hence we are able to generate an arbitrary three-level state (see Supplemental Material~\cite{SM} for more details). Indeed, it is much easier to prepare the two-level state vectors with high fidelity than three-level states. Thus, we generate a set of states given as $|\phi^T_k\rangle=\{|0\rangle, |1\rangle, |2\rangle, ( |0\rangle + |1\rangle)/\sqrt{2}, ( |0\rangle + |2\rangle)/\sqrt{2}, ( |1\rangle + |2\rangle)/\sqrt{2}, ( |0\rangle + \mathrm{e}^{-i2\pi /3} |1\rangle)/\sqrt{2}, ( |0\rangle + \mathrm{e}^{-i4\pi /3} |1\rangle)/\sqrt{2}, ( |0\rangle + \mathrm{e}^{-i8\pi /3} |2\rangle)/\sqrt{2},( |0\rangle + \mathrm{e}^{-i4\pi /3} |2\rangle)/\sqrt{2}, ( |1\rangle + \mathrm{e}^{-i2\pi /3} |2\rangle)/\sqrt{2}, ( |1\rangle + \mathrm{e}^{-i4\pi /3} |2\rangle)/\sqrt{2}\}$, rather than the vectors~(\ref{b1}) and (\ref{b2}) which can be decomposed into linear combinations of these $|\phi^T_k\rangle$~\cite{SM}. Finally, although it is impossible to implement a perfect BSM in linear optics~\cite{Cal2001}, the partial three-dimensional BSM $\{\mathcal{B}_1, \mathbb{I}-\mathcal{B}_1\}$ with $\mathcal{B}_1=|\Phi_3\rangle\langle\Phi_3|$ is possible to be realized. As displayed in Fig.~\ref{fig:1}, we pick the path $p_{11}, p_{22}$, and $p_{32}$ to encode the measurement projector $|\Phi_3\rangle$ and discard the remaining paths. Importantly, our method could be naturally applied to the trust-free verification of quantum steering  with $d\geq 3$ (see experimental details in Ref.~\cite{SM} for $d=4$).

{\it Results.---}As the first result, we report the measured parameter $S_{\rm EPR}$ as per Eq.~(\ref{eq:SI}), together with its theoretical expectation $S_{\rm EPR} (p)=2p+2\,(1-p)/3$ for the class of states~(\ref{mixedstate}). Our experimental data are marked with blue points and then fitted into the blue line in Fig.~\ref{fig:2}, while the theoretical prediction $S_{\rm EPR}(p)$ for $d=3$ is given as the red solid line and the bound $S_{\rm LHS}=1+1/\rt{3}$ for LHS models as the solid black line. The three blue points above the black straight line indicate that the experiment witnesses the violation of the steering inequality~(\ref{steerineq}), and thus we confirm the MDI verification of quantum steering for qutrits. Furthermore, we obtain $S_{\rm EPR}=1.983\pm0.002$ for $p=1$ from the fitted data, due to imperfections during the experiment. This bound $S(p=1)$, close to the quantum bound 2, implies that we have prepared the desired states with high fidelity in the sense that $p_{\rm eff}=0.987 p$~\cite{SM}. Additionally, the minimal $p$ for Alice to demonstrate steerability in our fitted line is $0.730$ while the theoretical one is $p_{\rm min}=0.683$; they are highlighted by the vertical dotted black lines respectively. By contrast, we also plot the theoretical predictions for $S_{\rm EPR}$ and the bound $S_{\rm LHS}$ for qubits in Fig.~\ref{fig:2}, and find that there is a noise-suppression phenomenon for high-dimensional EPR steering~\cite{Zeng2018}.

\begin{figure}[tbph]
	\includegraphics [width= 1.0\columnwidth]{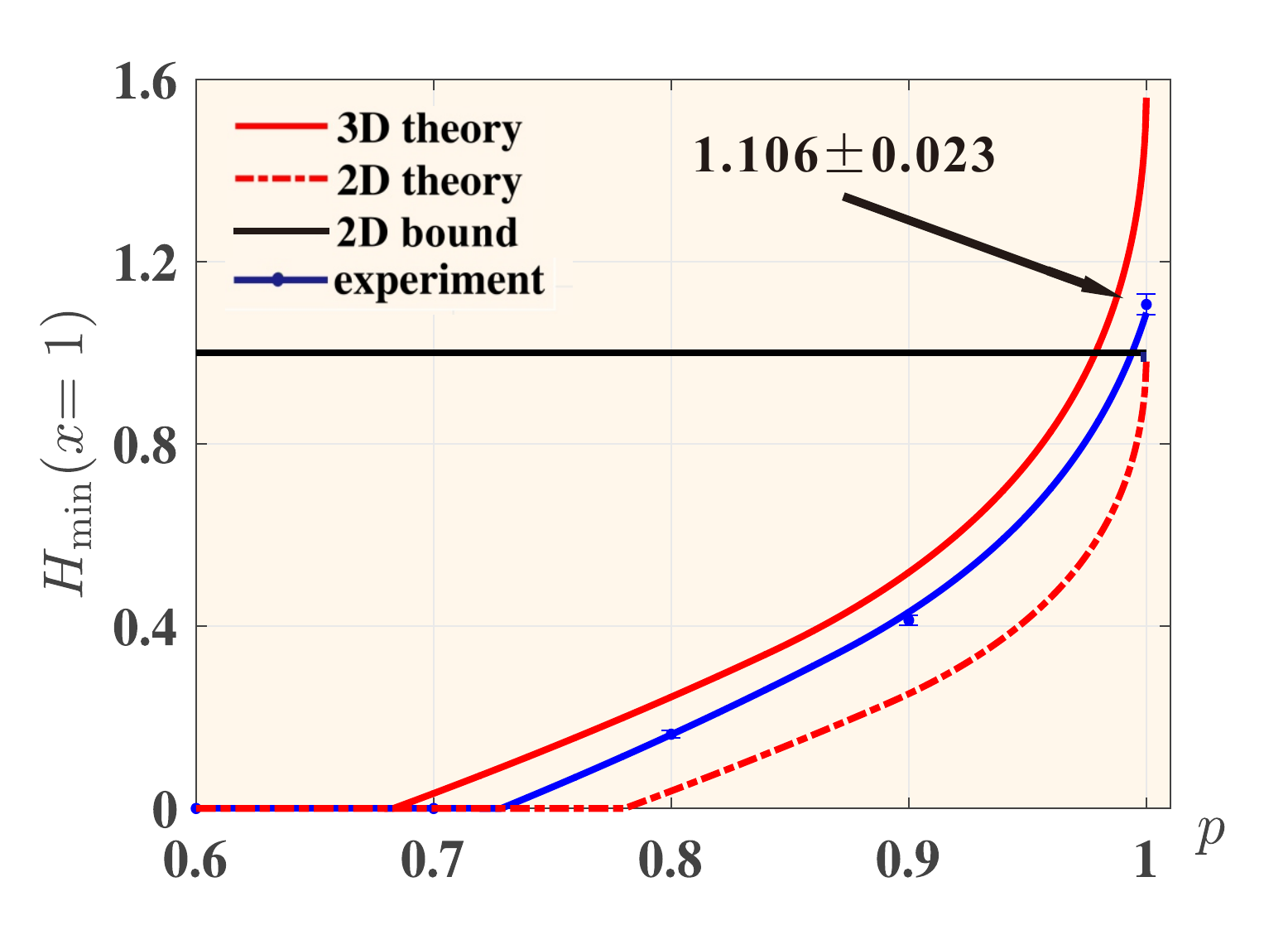}
	\caption{Randomness $H_{\rm min}$ with $p \in [0.6, 1]$. The blue dots describe the randomness generated from our observed data, and the solid blue curve corresponds to the fitting result, while the solid red curve is the theoretical prediction for states as per Eq.~(\ref{mixedstate}). By contrast, the expectation of randomness generation for qubit systems is plotted in the dotted red curve, and its maximal randomness with two settings is one bit (the horizontal black line). The error bars are of the order of $10^{-2}$.
	}
	\label{fig:3}
\end{figure}

Next, we apply our observed data to extract private randomness. As shown in Fig.~\ref{fig:3}, we investigate the amount of randomness $H_{\rm min}$ generated from Alice's measurements (for either setting $x$) with $p \in [0.6, 1]$ by using the semidefinite program from Ref.~\cite{Skrzypczyk2018}. It was proven in Ref.~\cite{Skrzypczyk2018} that the theoretical expectation for the three-dimensional case (solid red line) has an advantage over the two-dimensional one (dotted red line) in randomness generation, which is confirmed by our experimental results. The maximal randomness $H_{\rm min}(x^*)$ is achieved with $1.106\pm0.023$ bits per round, corresponding to the observed steering violation $S_{\rm EPR}=1.983\pm0.002$. It exceeds the one-bit bound for qubits (black line in Fig.~\ref{fig:3}) up to approximately 4.6 standard deviations. The result also implies that Charlie can certify randomness at a rate higher than consumed in choosing Alice's measurement setting and thus realizes a valid randomness expansion process. We also analyze the differences between experimentally observed $p(a, {\rm Yes}\,|j)$ in Eq.~(\ref{eq:SI}) and their theoretical predictions~\cite{SM}. The error bars of all the data are calculated from 100 simulations of Poisson statistics.

{\it Discussion.---}We have studied high-dimensional quantum steering, and experimentally demonstrated the trust-free verification of EPR steering beyond qubits by preparing a class of entangled photonic qutrits. We also applied our apparatus to extract randomness at a rate of $1.106\pm0.023$ per pair of photons, which exceeds the limit of the rate achievable by performing projective measurements on qubit systems and the amount of randomness consumed in choosing the measurement settings. To the best of our knowledge, this is the first valid MDI randomness expansion demonstration beyond qubits, which represents a significant step forward relative to the results reported in Ref.~\cite{Wang2018}. Our results can be generalized to higher-dimensional systems~\cite{SM}.

Another implication of our results is that since the maximal amount of randomness that can be certified with a pair of maximally entangled systems of dimension $d$ is $\mathrm{log}(d)$, then the certification of randomness of more than $\mathrm{log}(d)$ certifies genuine $d$-dimensional steering. So our results also certify genuine three-dimensional entanglement. This is consistent with recent results that measurements in two bases are sufficient for certifying high-dimensional entanglement~\cite{Bavaresco2018} and steering~\cite{Schneeloch2018}.

Regarding potential loopholes, they differ depending on the experimental goal. For demonstration of EPR steering, spacelike separation is essential, and could be overcome by using two remote particles~\cite{Pironio10, Yang18}; however for randomness generation this can be replaced by ensuring ``closed laboratory'' conditions. The MDI method can tolerate arbitrarily low detection efficiencies for demonstration of EPR-steering~\cite{pra2019}, however, it would in turn reduce the randomness certification rate. In addition, the randomness generation protocol in our experiment requires initial randomness (random choice of measurement settings in each experimental trial), which could be strengthened, e.g., by pseudorandomness based on the arrival time of cosmic photons~\cite{Wu17}. It is also of experimental interest to employ the randomness extraction process~\cite{Chung14, Shen18} to generate practical random bits.

\begin{acknowledgments} 
This Letter was supported by the National Key Research and Development Program of China (Grants No. 2017YFA0304100, No. 2016YFA0301300, and No. 2016YFA0301700), NSFC (Grants No. 11774335, No. 61327901, No. 11474268, No. 11504253, No. 11874345, and No. 11821404), the Key Research Program of Frontier Sciences, CAS (Grant No. QYZDY-SSW-SLH003), the Fundamental Research Funds for the Central Universities, and Anhui Initiative in Quantum Information Technologies (Grants Nos. AHY020100, and Nos. AHY060300). E. G. C. was supported by the Australian Research Council Centre of Excellence Project No. CE170100012 and Future Fellowship FT180100317. E. G. C. acknowledges useful discussions with Dr. Michael J. W. Hall.
\end{acknowledgments}

\bibliographystyle{unsrtnat}

\appendix

\widetext

\section{SUPPLEMENTAL INFORMATION}

\section{Note 1: Constructions of quantum-refereed steering witnesses}\label{note1}

We discuss how to construct a quantum-refereed steering witness (QRS) from a steering inequality. Start with an Einstein-Podolsky-Rosen (EPR)-steering inequality of the form:
\begin{eqnarray}
W_{\rm S} = \sum_j \an{a_j \hat{B}_j} \equiv \sum_{b,j} g_{b,j} \an{a_j E^B_{b|j}}=\sum_{a,b,j}g_{b,j}\,a\, p(a,b|j) \leq 0 \;, \label{SI:SI}
\end{eqnarray}
where each term is a correlation for $x=y=j$, and $\hat{B}_j \equiv \sum_b g_{b, j} E^B_{b|j}$ with $E^B_{b|j} \geq 0$ and $\sum_bE^B_{b|j}=\mathbb{I}$. When the measurement statistics $p(a, b|j)$ violates the inequality~(\ref{SI:SI}), it demonstrates the EPR steerability from Alice to Bob, or equivalently, the referee Charlie is convinced that they share entanglement.

However, in the measurement-device-independent (MDI) scenario, Charlie does not trust Bob to perform this positive-operator-valued measure (POVM) $\{E^B_{b|j}\}_{b, j}$ in the inequality~(\ref{SI:SI}). Indeed, instead of specifying Bob's measurement setting on a classical variable $j$, Charlie encodes it in a set of quantum states with density matrices $\{\tau^T_{b,j}\}_{b,j}$ on a Hilbert space $\mathcal{H}_C$ of dimension $d$ equal to that of $\mathcal{H}_B$, where $T$ is the transpose operation. Then, the most general thing Bob can do is to perform some arbitrary POVM $\{\mathcal{B}_i\}_i$ on $\mathcal{H}_B \otimes \mathcal{H}_C$, and answer ``Yes'' to Charlie's question when he obtains the outcome $\mathcal{B}_1$. Specifically, these question-states are chosen so that if Bob chooses to measure a POVM that includes the projector $\mathcal{B}_1 = | \Phi_d \rangle\langle \Phi_d |$ onto the maximally entangled state $| \Phi_d \rangle = \sum^{d-1}_{i=0} \frac{1}{\sqrt{d}} | i i\rangle$, then $E^B_{b|j}$ is proportional to the reduced POVM element ${\rm Tr}_C [(\mathbb{I}^B \otimes \tau^T_{b,j})\mathcal{B}_1]$ acting on Bob's system corresponding to this outcome. This can be done by choosing $E^B_{b|j}=\tau_{b,j}$, as we'll see in more detail below. In other words, sending $\tau^T_{b,j}$ to Bob corresponds to Charlie asking the question: ``If you were to perform measurement $j$, would you get the outcome $b$''?

Denote by $P(a,{\rm Yes}\,|j,\tau^T_{b,j})$ the probability that Alice answers $a$ to question $j$, and Bob answers ``Yes'' when he receives state $\tau^T_{b,j}$. Suppose now that Alice and Bob do not share a steerable state, i.e., suppose that there is a local hidden state (LHS) model as in Eq.~(1) in the main text. Then
\begin{align}
W_{\rm QRS}&=\sum_{a,b,j} g_{b,j} a_j P(a,{\rm Yes}\,|j,\tau^T_{b,j})  \nn \\
&= \sum_{b, j, \lambda} g_{b, j} p(\lambda) \an{ a_j}_\lambda \tr{(\rho^B_\lambda \otimes \tau^T_{b,j})\mathcal{B}_1}
= \sum_{b, j,\lambda} g_{b,j} p(\lambda) \an{a_j}_\lambda {\rm Tr}_C[\omega^T_\lambda \tau^T_{b,j}] \;,
\end{align}
where $\omega^T_\lambda \equiv {\rm Tr}_B[(\rho^B_\lambda \otimes \mathbb{I}^C)\mathcal{B}_1]=\frac{1}{d}(\rho^B_\lambda)^T$ are positive Hermitian operators acting on $\mathcal{H}^C$. Using ${\rm Tr}_C[\omega_\lambda \tau_{b,j}] = {\rm Tr}_C[\tau_{b,j}^T \omega^T_\lambda]$ $\tau_{b,j}=E^B_{b|j}$ and $\hat{B}_j \equiv \sum_b g_{b,j} E^B_{b|j}$, we are able to obtain
\begin{align}
W_{\rm QRS}&=\sum_{a,b,j} g_{b,j} a_j P(a,{\rm Yes}\,|j,\tau^T_{b,j})  \nn \\
&= \sum_{\lambda}  p(\lambda) \an{a_j}_\lambda {\rm Tr}_C[(\sum_{b,j} g_{b,j} \tau_{b,j})\omega_\lambda]= \frac{1}{d} \sum_{\lambda}  p(\lambda) \an{a_j}_\lambda {\rm Tr}_C[\hat{B}_j \rho^B_\lambda]  \nn \\
&= \frac{1}{d} W_S \leq 0 \; .
\end{align}

On the other hand, if Alice and Bob share an entangled state $\rho^{AB}$, and Alice measures POVMs $\{E^A_{a|j}\}_a$, and Bob measures a POVM with $\mathcal{B}_1 = | \Phi_d\rangle\langle \Phi_d |$, then we have
\begin{align}\label{eq:entangled}
W_{\rm QRS}&=\sum_{a,b,j} g_{b,j} a_j P(a,{\rm Yes}\,|j,\tau^T_{b,j})  \nn \\
&=\sum_{a,b,j} g_{b, j} a_j \tr{(E^A_{a|j} \otimes \mathcal{B}_1)(\rho^{AB} \otimes \tau^T_{b,j})} \nn \\
&=\frac{1}{d}\sum_{a,b,j} g_{b, j} a_j \tr{(E^A_{a|j} \otimes E^{B}_{b|j})\rho^{AB}}  \nn \\
&=\frac{1}{d}\sum_{j} \an{a_j \hat{B}_j} \; .
\end{align}
Note that the probability $P(a,{\rm Yes}\,|j,\tau^T_{b,j})$ in the QRS witness $W_{\rm QRS}$ and $p(a, b|j)$ in the steering inequality $W_{\rm S}$ obey an exact relation $P(a, {\rm Yes}\,|j,\tau^T_{b,j})=p(a, b|j)/d$.

\section{Note 2: Proof of the quantum steering inequality}\label{note2}

In this section, we give a rigorous proof to the steering inequality as per Eq.~(5) used in the main text
\begin{equation}
W_S=\sum_{a=b} p(a,b|1) + \sum_{a+b=0} p(a,b|2)-\left(1 + \frac{1}{\sqrt{d}}\right)\leq 0 \, ,
\end{equation}
where $a$ and $b$ are the outcomes of two mutually unbiased measurements $\hat{B}_j$, $j\in{1,2}$ on the d-dimensional system $B$, and the equality $a+b=0$ is the sum modulo $d$. When Alice and Bob share a non-steerable state, or, there is a LHS model to the data $p(a, b| j)$ as Eq.~(1), i. e.,
\beq
p(a,b|j)  =  \sum_\lambda p(\lambda)\, p(a|j,\lambda)\, \mathrm{Tr}[\Pi^B_{b|j} \rho^{B}_\lambda],
\eeq
where $\Pi^B_{b|j}\Pi_{b'|j}^B=\delta\, \Pi^B_{b|j}$, we are able to obtain
\begin{align}
S&=\sum_{a=b} p(a,b|1) + \sum_{a+b=0} p(a,b|2)\nn \\
&=\sum_\lambda p(\lambda)\left(\tr{(\sum_ap(a|1, \lambda)\Pi^B_{b=f_1(a)|1})\rho^B_\lambda}+\tr{(\sum_ap(a|2, \lambda)\Pi^B_{b=f_2(a)|2})\rho^B_\lambda}\right) \nn \\
&\equiv \sum_\lambda p(\lambda)\tr{(X_\lambda+Y_\lambda)\rho^B_\lambda},
\end{align}
where $p(a|j, \lambda)$ is the probobality distribution satisfying $\sum_a p(a| j, \lambda)=1, ~\forall j, \lambda$, and the functions are chosen to be $f_1(a)=a, f_2(a)+a=0~~{\rm mod}~~ d$. Moreover, it is easy to check that the positive operators $X_\lambda, Y_\lambda$ satisfy
\beq
\tr{X_\lambda}=\tr{Y_\lambda}=1,
\eeq
and then using $\mathrm{Tr}[\Pi_{b|j}\Pi_{b'|j'}]=1/d$ with $j\neq j'$ leads to
\beq
\tr{X_\lambda Y_\lambda}=\frac{1}{d}.
\eeq
First, it follows from the von Neumann trace inequality that
\beq
\tr{(X_\lambda+Y_\lambda)\rho^B_\lambda}\leq \sum_i s_i(X_\lambda+Y_\lambda)t_i(\rho^B_\lambda)\leq s_1(X_\lambda+Y_\lambda),
\eeq
where $s_i, t_i$ arranged in the decreasing order are the respective singular values of $X_\lambda+Y_\lambda$ and $\rho^B_\lambda$ , and the second inequality follows from that the density operator $\rho^B_\lambda$ has a maximal singular value 1 when it is a pure state. Then, if the $|\psi\rangle$ is the eigenvector corresponding to th eigenvalue $s_1$ of the matrix $X_\lambda+Y_\lambda$, we have
\begin{align}
s^2_1(X_\lambda+Y_\lambda)&=\langle \psi|X_\lambda+Y_\lambda|\psi\rangle   \nn \\
&=\sum_a \left(p(a|1,\lambda)\langle \psi|\Pi^B_{b=f_1(a)|j}|\psi\rangle+p(a|2,\lambda)\langle \psi|\Pi^B_{b=f_2(a)|j}|\psi\rangle\right) \nn \\
&= \max_a \left(\langle \psi|\Pi^B_{b=f_1(a)|j}|\psi\rangle+\langle \psi|\Pi^B_{b=f_2(a)|j}|\psi\rangle\right) \nn \\
&=(1+\frac{1}{\rt{d}})^2.
\end{align}
The third equality holds when the probobality distributions $p(a|j, \lambda)$ become deterministic distributions, i.e., there are some $p(a|j, \lambda)=1$ for the maximum $\langle\psi|\Pi^B_{b|j}|\psi\rangle$.  Finally, combining with above results immediately yields
\beq
S=\sum_\lambda p(\lambda)\tr{(X_\lambda+Y_\lambda)\rho^B_\lambda}\leq \sum_\lambda p(\lambda)\left(1+\frac{1}{\rt{d}}\right)=1+\frac{1}{\rt{d}},
\eeq
and we complete the proof of the steering inequality $W_{\rm S}\leq 0$.

For qutrits, i.e.,$d=3$, denote the eigenstates of $\hat{B}_1$ as
\begin{eqnarray}
|b_1=0\rangle & = & |0\rangle, \nn \\
|b_1=1\rangle & = & |1\rangle, \nn \\
|b_1=2\rangle & = & |2\rangle, \label{B1}
\end{eqnarray}
and we choose a MUB for the eingenstates of $\hat{B}_2$ to be
\begin{eqnarray}
|b_2=0\rangle & = & \frac{1}{\sqrt{3}} (|0\rangle + |1\rangle + |2\rangle), \nn \\
|b_2=1\rangle & = & \frac{1}{\sqrt{3}} (|0\rangle + \mathrm{e}^{i 2\pi/3}|1\rangle + \mathrm{e}^{i 4\pi/3} |2\rangle), \nn \\
|b_2=2\rangle & = & \frac{1}{\sqrt{3}} (|0\rangle + \mathrm{e}^{i 4\pi/3}|1\rangle + \mathrm{e}^{i 8\pi/3} |2\rangle). \label{B2}
\end{eqnarray}
With this choice of measurements, and with the maximally entangled state $|\Phi_3\rangle = \frac{1}{\sqrt{3}}\sum_{i=0}^2 |i\rangle |i\rangle$, the steering paramter $S$ achieves its maximal value 2 and thus there is the maximal quantum violation of the above steering inequality $W_{\rm S}=1-\frac{1}{\rt{3}}$.

\section{Note 3: Linear decompositions of the question-states $\{\tau^T_{b,j}\}$ sent from Charlie } \label{note3}

As discussed in the first section, we can write the quantum-mechanical prediction for each term $p(a, b|j)$ in the steering inequality~(\ref{SI:SI}) into the one $P(a, {\rm Yes}\,|j, \tau^T_{b,j})$ in the QRS witness~(\ref{eq:entangled}) as
\begin{align}
p(a,b|j)& = \tr{(E^A_{a|j} \otimes E^{B}_{b|j})\rho^{AB}}  \nn \\
&= d\, P(a, {\rm Yes}\,|j, \tau^T_{b,j})= d\, \tr{(E^A_{a|j} \otimes \mathcal{B}_1)(\rho^{AB} \otimes \tau^T_{b,j})}.
%&= d\, \tr{(E^A_{a|j} \otimes \mathcal{B}_1)(\rho^{AB} \otimes \sum_k s_{bjk} \tau^T_{k})}\nn \\
%&= d\, \sum_k s_{bjk} \tr{(E^A_{a|j} \otimes \mathcal{B}_1)(\rho^{AB} \otimes \tau^T_{k})}.
\end{align}
Note that the question-states $\{\tau^T_{b, j}\}$ sent from Charlie are chosen so that $E_{b|j}^B=\tau_{b, j}$. It immediately leads to the fact that the input states are given by the eigenstate vectors as per Eqs.~(\ref{B1}) and (\ref{B2}).  In this experiment, instead of preparing these 3-level states directly, we generate a set of 2-level state vectors $\{\tau^T_k\}$ which are much easier to be prepared with high fidelity. Indeed, we choose for the question-states $\tau_{k}^T=|\phi_{k}\rangle\langle\phi_{k}|^T$ with
\begin{flalign}
&|\phi_1\rangle=|0\rangle, \nn \\
&|\phi_2\rangle=|1\rangle, \nn\\
&|\phi_3\rangle=|2\rangle, \nn \\
&|\phi_4\rangle=\frac{1}{\sqrt{2}}( |0\rangle + |1\rangle), \nn \\
&|\phi_5\rangle=\frac{1}{\sqrt{2}}( |0\rangle + |2\rangle), \nn \\
&|\phi_6\rangle =\frac{1}{\sqrt{2}}( |1\rangle + |2\rangle), \nn \\
&|\phi_7\rangle =\frac{1}{\sqrt{2}}( |0\rangle + \mathrm{e}^{-i2\pi /3} |1\rangle), \nn \\
&|\phi_8\rangle =\frac{1}{\sqrt{2}}( |0\rangle + \mathrm{e}^{-i4\pi /3} |1\rangle), \nn \\
&|\phi_9\rangle =\frac{1}{\sqrt{2}}( |0\rangle + \mathrm{e}^{-i8\pi /3} |2\rangle), \nn \\
&|\phi_{10}\rangle =\frac{1}{\sqrt{2}}( |0\rangle + \mathrm{e}^{-i4\pi /3} |2\rangle), \nn \\
&|\phi_{11}\rangle =\frac{1}{\sqrt{2}}( |1\rangle + \mathrm{e}^{-i2\pi /3} |2\rangle), \nn \\
&|\phi_{12}\rangle =\frac{1}{\sqrt{2}}( |1\rangle + \mathrm{e}^{-i4\pi /3} |2\rangle). \label{input}
\end{flalign}

With this choice, it is easy to check that all $E_{b|j}s$ or $\tau_{b, j}s$ can be decomposed as linear combinations of $\{\tau_k^T\}$ as
\begin{flalign}
E_{0|1}=&|b_1=0\rangle\langle b_1=0|=\tau_{1}, \nn \\
E_{1|1}=&|b_1=1\rangle\langle b_1=1|=\tau_{2}, \nn \\
E_{2|1}=&|b_1=2\rangle\langle b_1=2|=\tau_{3}, \nn \\
E_{0|2}=&|b_2=0\rangle\langle b_2=0|=\frac{1}{3}(-\tau_{1}-\tau_{2}-\tau_{3}+2\tau_{4}+2\tau_{5}+2\tau_{6}), \nn \\
E_{1|2}=&|b_2=1\rangle\langle b_2=1|=\frac{1}{3}(-\tau_{1}-\tau_{2}-\tau_{3}+2\tau_{7}+2\tau_{10}+2\tau_{11}),  \nn \\
E_{2|2}=&|b_2=2\rangle\langle b_2=2|=\frac{1}{3}(-\tau_{1}-\tau_{2}-\tau_{3}+2\tau_{8}+2\tau_{9}+2\tau_{12}).
\end{flalign}
Further, we are able to obtain
\begin{align}
p(a,b|j)& = d\, P(a, {\rm Yes}\,|j, \tau^T_{b,j})= d\, \tr{(E^A_{a|j} \otimes \mathcal{B}_1)(\rho^{AB} \otimes \tau^T_{b,j})}, \nn \\
&= d\, \tr{(E^A_{a|j} \otimes \mathcal{B}_1)(\rho^{AB} \otimes \sum_k s_{bjk} \tau^T_{k})}\nn \\
&= d\, \sum_k s_{bjk} \tr{(E^A_{a|j} \otimes \mathcal{B}_1)(\rho^{AB} \otimes \tau^T_{k})},
\end{align}
where $E_{b|j}=\tau_{b, j}=\sum_ks_{bjk}\tau_k.$

\section{Note 4: Experimental implementation of Alice's measurements}

%%%%%%%%%%%%%%%%%%%%%%%%%%%%%%%%%%%%%%%%%%%%%%%%%%%%%%%%%%%%%%%%%%%%%%%%%%%%%%%%
%\begin{figure*}[htbp]
%[tbph]
%\begin{center}
%\includegraphics [width= 0.6\columnwidth]{alicesetting}
%\end{center}
%\caption{Realisation of Alice's measurements. H: half-wave plate; Q: quarter-wave plate; D: single photon detector.
%}
%\label{fig:1}
%\end{figure*}
%%%%%%%%%%%%%%%%%%%%%%%%%%%%%%%%%%%%%%%%%%%%%%%%%%%%%%%%%%%%%%%%%%%%%%%%%%%%%%%%

For the steering inequality~(5) in the main text, there are two measurement settings $x=1, 2$ for Alice. For the 3-dimensional system, these quantum measurements are chosen as the same as Bob's measurements, i.e., three measurement outcomes of the setting $x=1$ are given by Eq.~(\ref{B1}) while the setting $x=2$ has a description in Eq.~(\ref{B2}). As mentioned in the main text, Alice's measurement setting is realised via placing 5 half-wave plates (HWPs), a quarter-wave plate (QWP), two beam displacers (BDs), a polarising beam splitter (PBS), and 3 single photon detectors. In the steering scenario, Alice can randomly choose one of the measurement settings, which can be done by rotating the angles of HWPs properly. Specifically, tuning the QWP at $0^\circ$, we set HWP1-5 at $45^\circ$, $0^\circ$, $45^\circ$, $45^\circ$, and $0^\circ$ for the setting $x=1$, and set HWP1-5 at $45^\circ$, $67.5^\circ$, $72.37^\circ$, $45^\circ$, and $22.5^\circ$ for the setting $x=2$. The detectors D1-D3 are used to record three outcomes 0-2.

\section{Note 5: Preparation of the trusted input states}

%%%%%%%%%%%%%%%%%%%%%%%%%%%%%%%%%%%%%%%%%%%%%%%%%%%%%%%%%%%%%%%%%%%%%%%%%%%%%%%%
\begin{figure*}[htbp]
%[tbph]
\begin{center}
\includegraphics [width= 0.6\columnwidth]{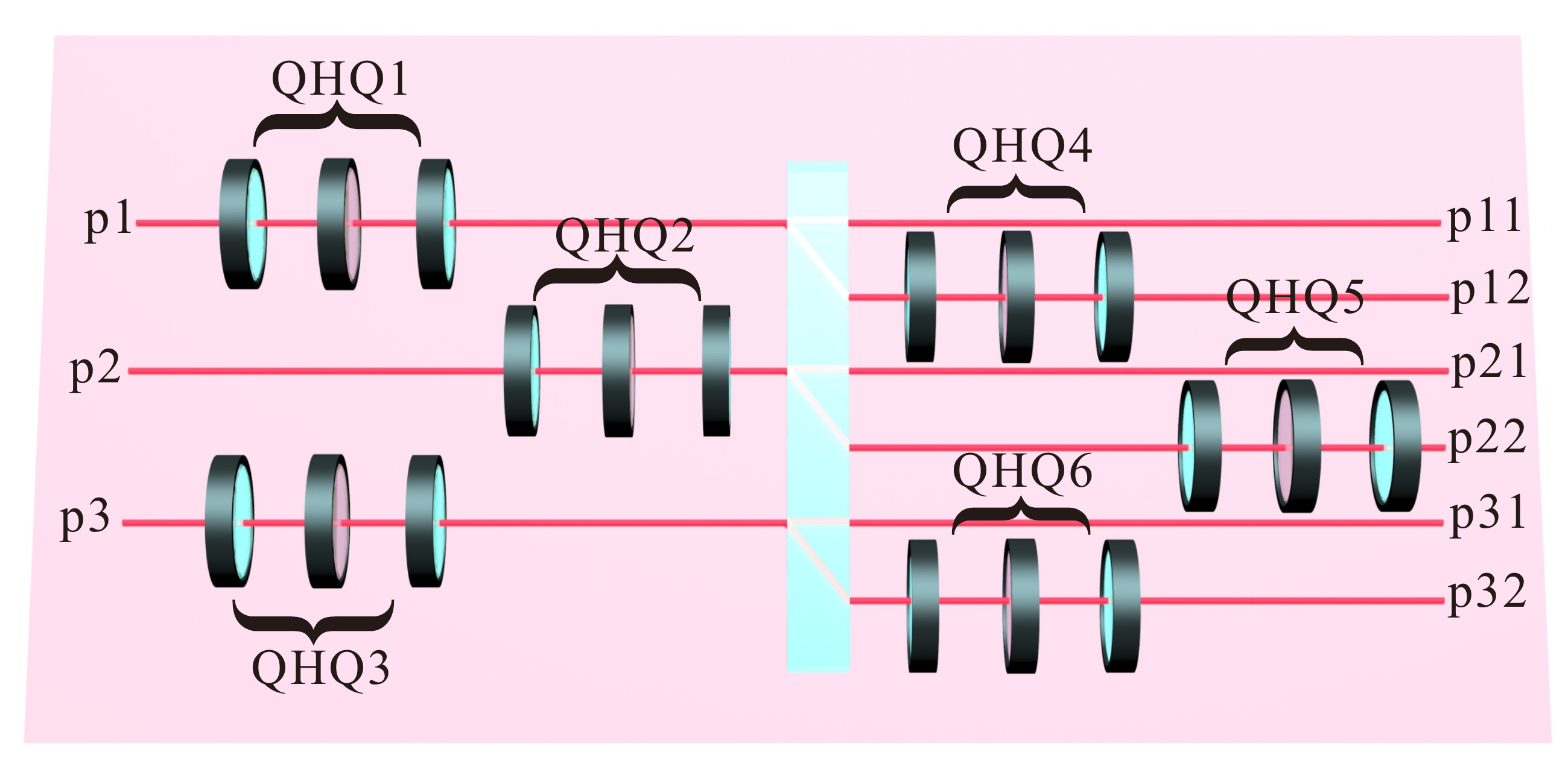}
\end{center}
\caption{Experimental preparation of the question-states $\{\tau^T_k\}$. H: half-wave plate; Q: quarter-wave plate.
}
\label{fig:2}
\end{figure*}
%%%%%%%%%%%%%%%%%%%%%%%%%%%%%%%%%%%%%%%%%%%%%%%%%%%%%%%%%%%%%%%%%%%%%%%%%%%%%%%%

As plotted in Fig.~\ref{fig:2}, these question-states are prepared by passing 6 wave plate assemblages, each of which contains two QWPs and a HWP sandwiched between them (QHQ assemblage), and encoded in the path- and polarisation- degrees of freedom of photons. In particular, the photon passing the path p1 encodes the state $|0\rangle$, the photon in the path p2 is the state $|1\rangle$, and photons in the path p2 are encoded as the state $|2\rangle$. Moreover, the vertically polarised (V) photons in the path p11 (p21, p31) describe the state $|\tilde{0}\rangle$, horizontally polarised (H) photons in the path p12 (p22, p32) encode the state $|\tilde{1}\rangle$, and V photons in the path p12 (p22, p32) encode the state $|\tilde{2}\rangle$. Since the QHQ assemblage is constructed to realise an arbitrary unitary operation on qubit states, QHQ1 and QHQ4 would transform the state $|0\rangle$ into $|0\rangle\otimes\left(\beta_0|\tilde{0}\rangle+\beta_1|\tilde{1}\rangle+\beta_2|\tilde{2}\rangle\right)$, where $\beta_i$ are complex coefficients. Thus, if QHQ2 and QHQ3 (QHQ5 and QHQ6) are synchronized with QHQ1 (QHQ4), then an arbitrary state $\alpha_0|0\rangle+\alpha_1|1\rangle+\alpha_2|2\rangle$ can be transformed to a state $(\alpha_0|0\rangle+\alpha_1|1\rangle+\alpha_2|2\rangle)\otimes(\beta_0|\tilde{0}\rangle+
\beta_1|\tilde{1}\rangle+\beta_2|\tilde{2}\rangle)$. By choosing $\beta_s$ properly, we are able to prepare the states $\{\tau_k^T\}$ with high fildelity.

\section{Note 6: Implementation of the partial 3-dimensional Bell state measurement}

%%%%%%%%%%%%%%%%%%%%%%%%%%%%%%%%%%%%%%%%%%%%%%%%%%%%%%%%%%%%%%%%%%%%%%%%%%%%%%%%
\begin{figure*}[htbp]
%[tbph]
\begin{center}
\includegraphics [width= 0.6\columnwidth]{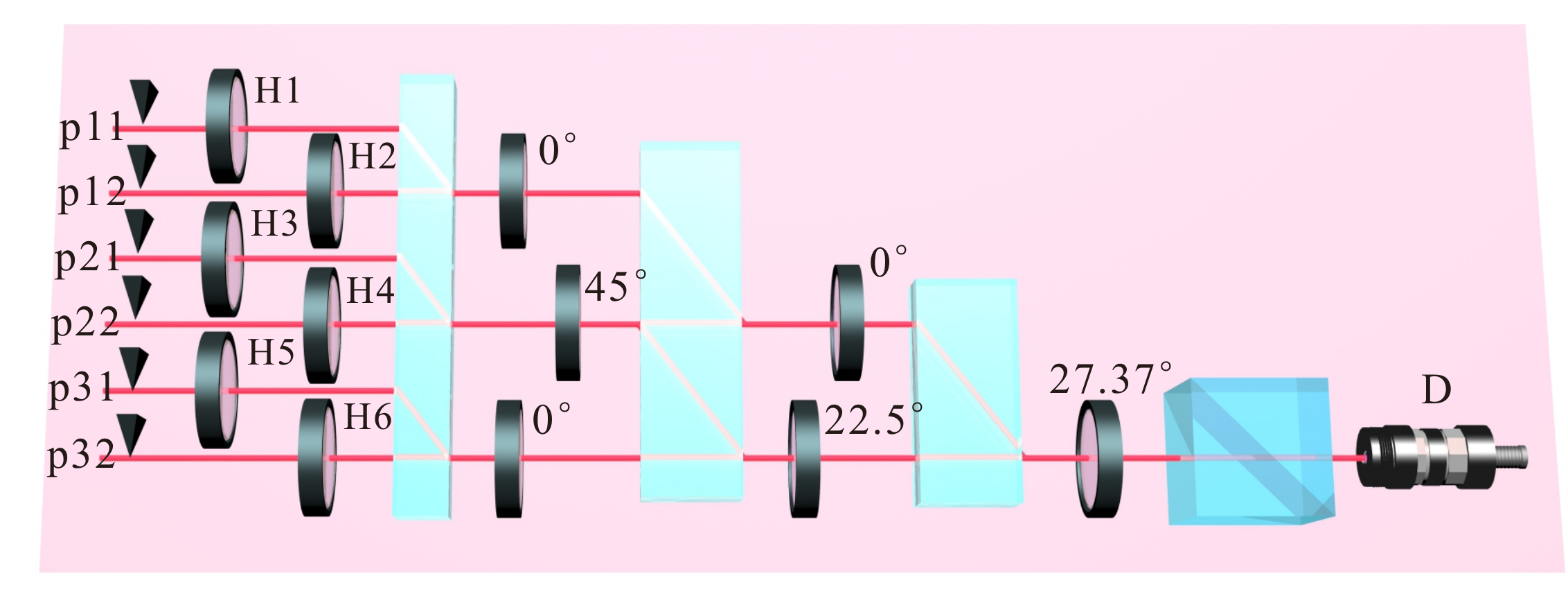}
\end{center}
\caption{Realisation of the measurement projector $\mathcal{B}_1=|\Phi_3\rangle\langle\Phi_3|$. H: half-wave plate; D: single photon detector.
}
\label{fig:3}
\end{figure*}
%%%%%%%%%%%%%%%%%%%%%%%%%%%%%%%%%%%%%%%%%%%%%%%%%%%%%%%%%%%%%%%%%%%%%%%%%%%%%%%%

There are 9 Bell states for the two-qutrit systems, however, we just need to consider the partial BSM $\{\mathcal{B}_1, \mathbb{I}-\mathcal{B}_1\}$ that includes the unique projector $\mathcal{B}_1=|\Phi_3\rangle\langle\Phi_3|$ on the maximally entangled state
\begin{eqnarray}
|\Phi_{3}\rangle & = & \frac{1}{\sqrt{3}} (|00\rangle + |11\rangle + |22\rangle).
\end{eqnarray}
This partial BSM is much easy to implement experimentally as we only need to generate the measurement projector $\mathcal{B}_1$. For the measurement vector $|\Phi_{3}\rangle$, we block the paths p12, p21, and p31, and set HWP1, HWP4, and HWP6 to $0^\circ$, $45^\circ$, and $0^\circ$, while the angles of the rest HWPs are specified in Fig.~\ref{fig:3}.

\section{Note 7: Noise analysis of the experimental data}

%%%%%%%%%%%%%%%%%%%%%%%%%%%%%%%%%%%%%%%%%%%%%%%%%%%%%%%%%%%%%%%%%%%%%%%%%%%%%%%%
\begin{figure*}[htbp]
	%[tbph]
	\begin{center}
		\includegraphics [width= 0.6\columnwidth]{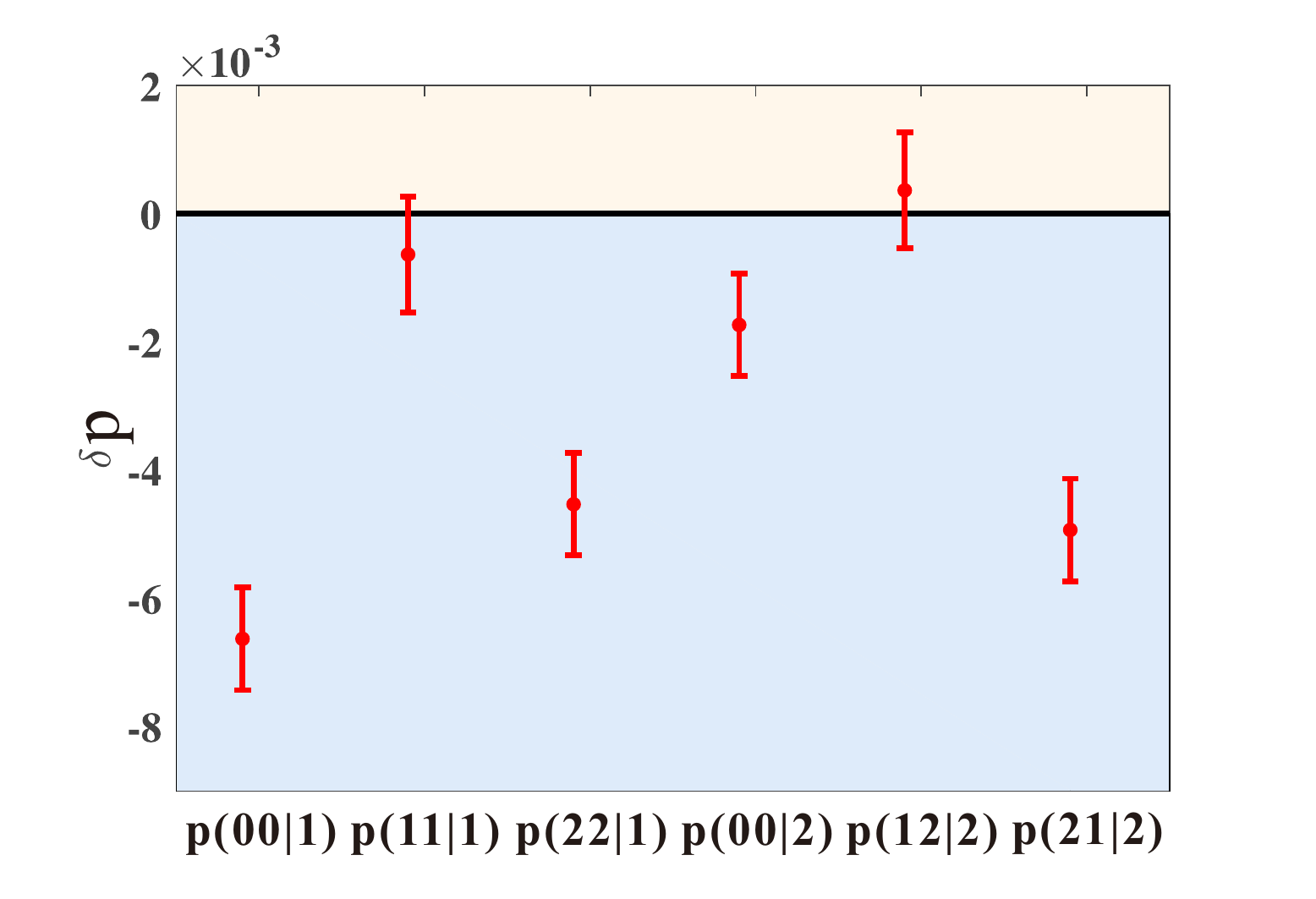}
	\end{center}
	\caption{The experimental data $p(a, b|j)$ are compared with their theoretical predictions. The error bars of all the data are calculated from a standard deviation of 100 simulations of Poisson statistics.
	}
	\label{fig:7}
\end{figure*}
%%%%%%%%%%%%%%%%%%%%%%%%%%%%%%%%%%%%%%%%%%%%%%%%%%%%%%%%%%%%%%%%%%%%%%%%%%%%%%%%

In Fig.~\ref{fig:7}, we compare the measurement statistics $p(a, b|j)$ in Eq.~(4) in the main text with their theoretical values, i.e., $p=p_{\rm exp}-p_{\rm theo}$, when Alice and Bob share the maximally entamgled state $|\Phi_3\rangle$. We also find that the points in the light blue area decrease the experimental value of $S_{\rm EPR}$ and lowers the randomness $H_{\rm min}(x=1)$ either.

\section{Note 8: Schemes for 4-dimensional MDI steering}

In the 4-dimensional MDI steering protocol, the steering inequality derived in the main text can be adapted as
\begin{equation}\label{eq:SI}
S_{\rm EPR}=\sum_{x=1}^2 \sum_{a,b=f_x(a)} P(a,b|x) \leq S_{\rm LHS}= \frac{3}{2} \, .
\end{equation}
Similar to the qutrit case, the maximally entangled qudit state $|\Phi_4\rangle=1/2\sum_{i=0}^3|ii\rangle$ can be prepared via the hybrid of the path and polarisation source with high fidelity~\cite{Hu2018,Guo18}, and the isotropic noise can also be added to prepare a class of states $\rho=p|\Phi_4\rangle\langle\Phi_4|+\frac{1-p}{16}I$. Then, what is left to do is simulate Alice's measurements, Bob's question-states sent from Charlie, and the partial BSM on Bob's subsytem and the received states, all of which are acting on the 4-dimensional Hilbert state space. We'll briefly discuss the experimetal details to perform the trust-free verification of 4-dimensional quantum steering.

\subsection{A: Alice's measurements}
%%%%%%%%%%%%%%%%%%%%%%%%%%%%%%%%%%%%%%%%%%%%%%%%%%%%%%%%%%%%%%%%%%%%%%%%%%%%%%%%
\begin{figure*}[htbp]
%[tbph]
\begin{center}
\includegraphics [width= 0.6\columnwidth]{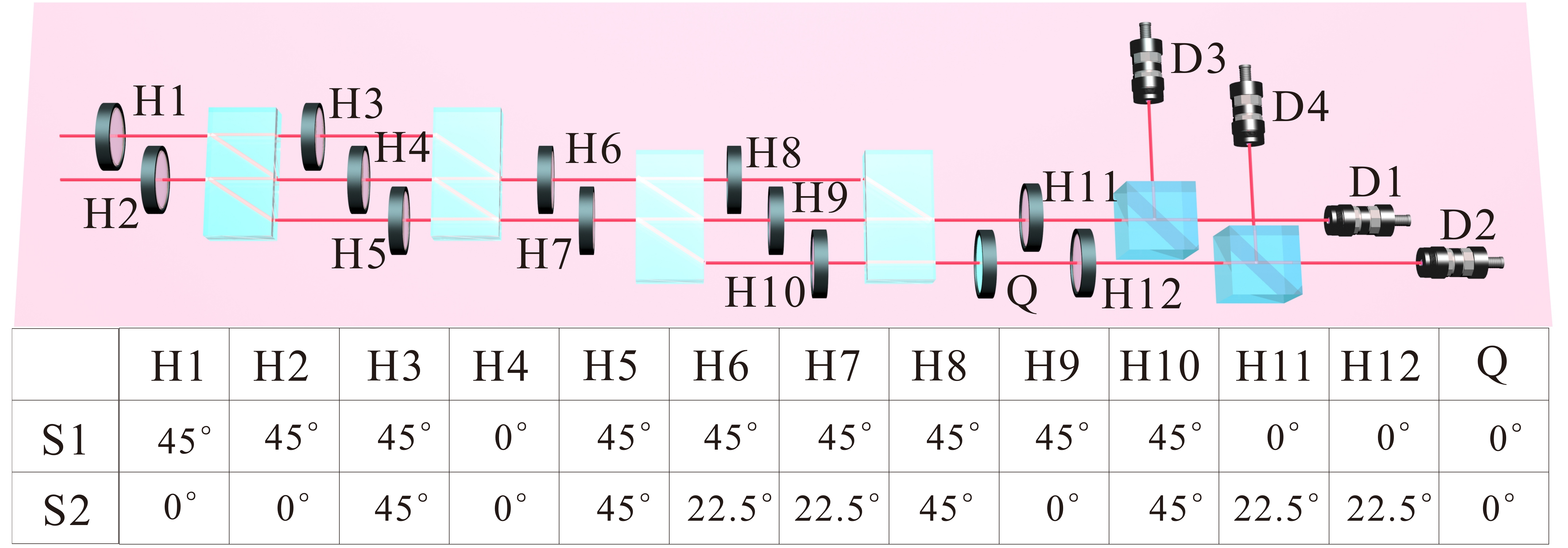}
\end{center}
\caption{Simulation of Alice's measurements in the 4-dimensional case. H: half-wave plate; Q: quarter-wave plate; D: single photon detector.
}
\label{fig:4}
\end{figure*}
%%%%%%%%%%%%%%%%%%%%%%%%%%%%%%%%%%%%%%%%%%%%%%%%%%%%%%%%%%%%%%%%%%%%%%%%%%%%%%%%

For the steering inequality Eq.~(\ref{eq:SI}), the eigenstates, corresponding to measurement outcomes of the setting $x=1$, have a quantum description as
\begin{eqnarray}
|b_1=0\rangle & = & |0\rangle, \nn \\
|b_1=1\rangle & = & |1\rangle, \nn \\
|b_1=2\rangle & = & |2\rangle, \nn  \\
|b_1=3\rangle & = & |3\rangle.
\end{eqnarray}
And for the setting $x=2$, there are
\begin{eqnarray}
|b_2=0\rangle & = & \frac{1}{2} (|0\rangle + |1\rangle + |2\rangle + |3\rangle), \nn \\
|b_2=1\rangle & = & \frac{1}{2} (|0\rangle + \mathrm{e}^{i \pi/2}|1\rangle + \mathrm{e}^{i \pi} |2\rangle + \mathrm{e}^{i 3\pi/2}|3\rangle), \nn \\
|b_2=2\rangle & = & \frac{1}{2} (|0\rangle + \mathrm{e}^{i \pi}|1\rangle + \mathrm{e}^{i 2\pi} |2\rangle + \mathrm{e}^{i 3\pi}|3\rangle), \nn \\
|b_2=3\rangle & = & \frac{1}{2} (|0\rangle + \mathrm{e}^{i 3\pi/2}|1\rangle + \mathrm{e}^{i 3\pi}|2\rangle + \mathrm{e}^{i 9\pi/2}|3\rangle).
\end{eqnarray}

Similarly, we can use basic optical elements, including HWPs and QWPs, to simulate these two measurement settings, which is shown in Fig.~\ref{fig:4}. All necessary parameters related to the optical elements are given  in the table embedded in Fig.~\ref{fig:4} (the second row for the setting $x=1$ and the third row for the setting $x=2$).

\subsection{B: Bob's received states}
%%%%%%%%%%%%%%%%%%%%%%%%%%%%%%%%%%%%%%%%%%%%%%%%%%%%%%%%%%%%%%%%%%%%%%%%%%%%%%%%
\begin{figure*}[htbp]
%[tbph]
\begin{center}
\includegraphics [width= 0.6\columnwidth]{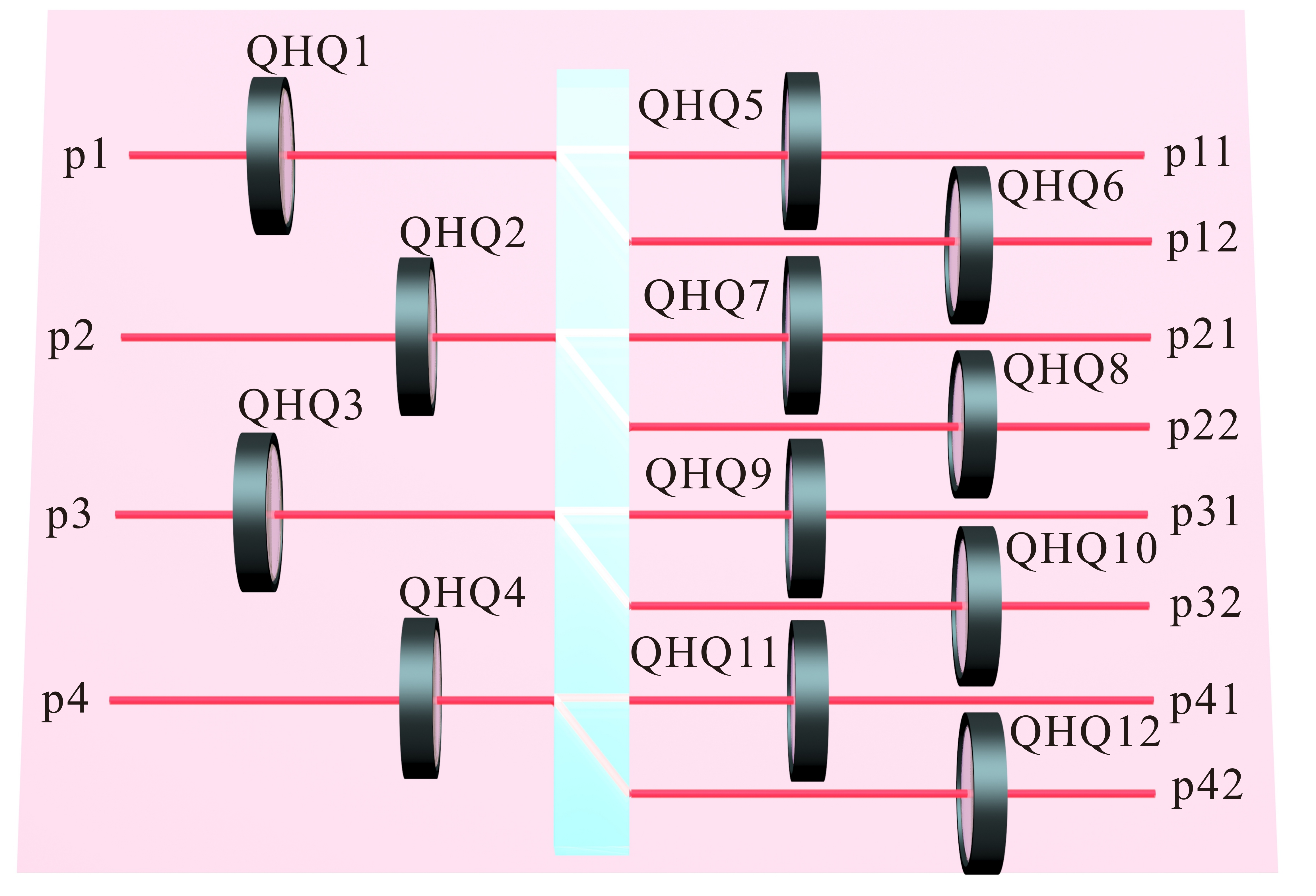}
\end{center}
\caption{Experiment implementation of 4-dimensional question-states. QHQ: a wave plate assemblage which is made of 2 QWPs and a HWP.
}
\label{fig:5}
\end{figure*}
%%%%%%%%%%%%%%%%%%%%%%%%%%%%%%%%%%%%%%%%%%%%%%%%%%%%%%%%%%%%%%%%%%%%%%%%%%%%%%%%
It is shown in Fig.~\ref{fig:5} that the 4-dimensional question-states send from Charlie can be faithfully generated. In particular, the photons going through the path p1 encode the state $|0\rangle$, photons in the path p2 encode the state $|1\rangle$, photons in the path p2 encode the state $|3\rangle$, and the photons passing the path p4 encode the state $|3\rangle$. On the right side of the wave plate,  H photons in the path p11 (p21, p31, p41) are encoded as the state $|\tilde{0}\rangle$, V photons in the path p11 (p21, p31, p41) as $|\tilde{1}\rangle$, H photons in the path p12 (p22, p32, p42) as $|\tilde{2}\rangle$, and V photons in the path p12 (p22, p32, p42) as $|\tilde{3}\rangle$. The wave plate assembalges QHQ1, QHQ5, and QHQ6 are tuned to prepare the state $|0\rangle$ into $|0\rangle\otimes(\beta_0|\tilde{0}\rangle+\beta_1|\tilde{1}\rangle+
\beta_2|\tilde{2}\rangle+\beta_3|\tilde{3}\rangle)$. Then, synchronizing QHQ2, QHQ3, and QHQ4 (QHQ7, QHQ9, QHQ11 and QHQ8, QHQ10, QHQ12) with QHQ1 (QHQ5 and QHQ6), we can prepare an arbitrary state $\alpha_0|0\rangle+\alpha_1|1\rangle+\alpha_2|2\rangle+\alpha_3|3\rangle$ to the state $(\alpha_0|0\rangle+\alpha_1|1\rangle+\alpha_2|2\rangle+\alpha_3|3\rangle)\otimes(\beta_0|\tilde{0}\rangle+
\beta_1|\tilde{1}\rangle+\beta_2|\tilde{2}\rangle+\beta_3|\tilde{3}\rangle)$. Finally, choosing these complex coefficients $\beta_s$ properly yields an arbitrary 4-dimensional pure input states.

\subsection{C: Partial 4-dimensional BSM}

%%%%%%%%%%%%%%%%%%%%%%%%%%%%%%%%%%%%%%%%%%%%%%%%%%%%%%%%%%%%%%%%%%%%%%%%%%%%%%%%
\begin{figure*}[htbp]
%[tbph]
\begin{center}
\includegraphics [width= 0.6\columnwidth]{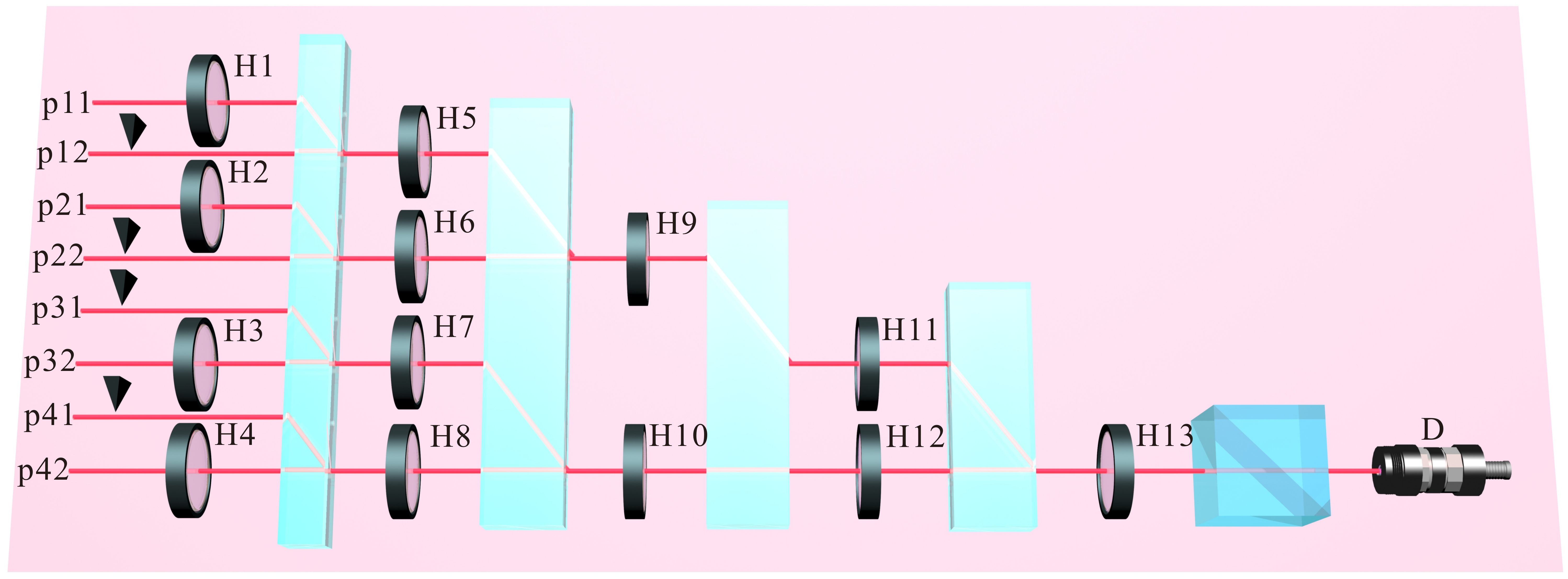}
\end{center}
\caption{Implementation of the measurement projector $|\Phi_4\rangle\langle\Phi_4|$. H: half-wave plate; D: single photon detector.
}
\label{fig:6}
\end{figure*}
%%%%%%%%%%%%%%%%%%%%%%%%%%%%%%%%%%%%%%%%%%%%%%%%%%%%%%%%%%%%%%%%%%%%%%%%%%%%%%%%

The partial 4-dimensional Bell state measurement $\{\mathcal{B}_1, \mathbb{I}-\mathcal{B}_1\}$ with $\mathcal{B}_1=|\Phi_{4}\rangle\langle\Phi_4|$ can be implemented in the experiment in a similar way as  that of qutrit systems. Indeed, to construct the projective operator of $|\Phi_{4}\rangle = \frac{1}{2}(|00\rangle + |11\rangle + |22\rangle + |33\rangle)$, we just need to keep H photons in p11, V photons in the path p21, H photons in the p32, and V photons in the p42 and discard the rest paths (see Fig.~\ref{fig:6}). To be specific, the output of the PBS in the BSM module corresponds to $|\Phi_{4}\rangle$ when the HWP1-13 are set to the angles at $0^\circ$, $45^\circ$, $45^\circ$, $0^\circ$, $0^\circ, 45^\circ$, $45^\circ$, $0^\circ$, $22.5^\circ$, $22.5^\circ, 0^\circ$, $0^\circ$, and $22.5^\circ$.

\end{document}